\documentclass[onecolumn,secnumarabic,amssymb, nobibnotes,aps,preprint,pra]{revtex4-1}
\usepackage{color}
\usepackage{amsmath,amssymb}
\usepackage{bm}
\usepackage{graphicx}

\begin{document}

\title{Near-field cavity optomechanical coupling in a compound semiconductor nanowire}%

\author{Motoki Asano}%
\author{Guoqiang Zhang}
\author{Takehiko Tawara}
\author{Hiroshi Yamaguchi}
\author{Hajime Okamoto}

\affiliation{NTT Basic Research Laboratories, NTT Corporation, 3-1 Morinosato Wakamiya, Atsugi-shi, Kanagawa 243-0198, Japan}

\begin{abstract}{\bf
A III-V compound semiconductor nanowire is an attractive material for a novel hybrid quantum interface that interconnects photons, electrons, and phonons through a wavelength-tunable quantum structure embedded in its free-standing structure. In such a nanomechanical element, however, a challenge is how to detect and manipulate a small number of phonons via its tiny mechanical motion. A solution would be to couple an optical cavity to a nanowire by introducing the ``cavity optomechanics'' framework, but the typical size difference between them becomes a barrier to achieving this. Here, we demonstrate near-field coupling of a silica microsphere cavity and an epitaxially grown InP/InAs free-standing nanowire. The evanescent optomechanical coupling enables not only fine probing of the mechanical motion by balanced homodyne interferometry but also tuning of the resonance frequency, linewidth, Duffing nonlinearity, and vibration axis in the nanowire. Combining this cavity optomechanics with epitaxial nanowire engineering opens the way to novel quantum metrology and information processing.
}
\end{abstract}
\date{\today}%
\maketitle

\section*{Introduction}
Semiconductor nanowires are key elements in not only nanophotonics \cite{yan2009nanowire} and nanoelectronics \cite{thelander2006nanowire} but also in nanomechanics owing to their rich mechanical properties (e.g. small-effective mass and high aspect ratio) \cite{feng2007very}. A III-V compound semiconductor nanowire is a strong candidate for a wavelength-tunable hybrid quantum interface because the resonant photon energy of the built-in semiconductor low-dimensional (quantum) structure can be widely tuned from visible to telecom wavelengths by epitaxial engineering \cite{saxena2016design,zhang2019telecom}. Moreover, coupling between mechanical motion, namely phonons, and quantum excitonic states via strain induced by the deformation potential and piezoelectric effect has been demonstrated in III-V compound semiconductor nanowires \cite{montinaro2014quantum,yeo2014strain,munsch2017resonant}. Although bringing the nanowire mechanical motion into a few-phonon regime (especially into a single phonon) can dramatically promote the functionality in hybrid quantum interfaces \cite{stannigel2011optomechanical,restrepo2017fully}, Precisely detecting and controlling such a tiny mechanical motion by electric or optical means are technically difficult because of the nanowire's small electrostatic capacity, low optical absorbance, and reflectivity.

One way to probe and control a few phonons in a free-standing nanowire is to introduce a framework called ``cavity optomechanics'' \cite{aspelmeyer2014cavity} by coupling a high-Q optical (photon) cavity to a nanowire mechanical (phonon) resonator to induce photon-phonon interactions. However, such cavity coupling is quite challenging because of the typical size difference between the optical cavity and nanowires. Therefore, very few demonstrations of cavity optomechanics in nanowire system have been reported \cite{fogliano2019cavity}, and implementing it in III-V compound semiconductors has not yet been attempted. Applying cavity optomechanics in III-V compound semiconductor nanowires would lead to rich optomechanical operation in the wide range of optical wavelengths as well as to hybrid quantum optoelectromechanics through wavelength-tunable quantum structures.

Here, we demonstrate near-field optomechanical coupling of a movable silica whispering-gallery mode (WGM) microsphere cavity (Q = $1.8 \times 10^5$) and an InP/InAs free-standing nanowire through an evanescent gradient force field. This near-field approach allows us to achieve a free access of an optical cavity to a subwavelength-scale mechanical element such as a free-standing semiconductor nanowire based on the use of fiber optics \cite{anetsberger2009near,cole2015evanescent,brawley2016nonlinear,schilling2016near,asano2018opto,magrini2018near}. It enables us to not only finely probe the thermal motion of a single nanowire with sub-picometer sensitivity by balanced homodyne interferometry but also to tune the frequency, linewidth, Duffing nonlinearity, and vibration axis of the two orthogonalized mechanical modes. Our results show that, by further optimizing the fiber-cavity positioning apparatus, quantum-limited displacement detection is available in this nanowire system. This approach can be extended to epitaxially engineered nanowires in a cryogenic environment in which a semiconductor low-dimensional (quantum) structure resonantly couples to external cavity photons, which would open the way to the development of a novel hybrid quantum nanomechanical systems.

The concept of this study is illustrated in Fig. 1a. In our system, the optomechanical coupling is in the bad-cavity regime, where the optical linewidth ($\kappa$) is larger than the mechanical mode frequency ($\Omega$) \cite{aspelmeyer2014cavity}. In such a regime, the coupling enables high-sensitivity detection of mechanical motion by monitoring the optical transmission, whose phase is modulated with the mechanical mode frequency via the optomechanical coupling (see Fig. 1b). On the other hand, the optical force causes the modulation of linear mechanical responses (frequency shifts and linewidth broadening) (see Fig. 1c). In particular, the on-resonance laser light dominantly leads to a static optical force regarded as quadratic optomechanical coupling. This force results in a rotation of the vibration axis of the two orthogonalized mechanical modes. In the following, we demonstrate high-sensitive displacement detection and optical control of mechanical motion in a single nanowire using this near-field cavity optomechanical coupling.

\section*{Results}
\subsection*{High-sensitivity displacement measurement}

The near-field cavity optomechanical system was formed with an epitaxially grown free-standing nanowire and a movable WGM microsphere. Nanowires containing InP/InAs heterostructures in the vertical (growth) direction were epitaxially grown on an InP substrate by metal organic vapor phase epitaxy (see Methods). The length and diameter of the nanowires are about 14 $\mathrm{\mu m}$ and 500 nm, respectively, which was determined from the scanning electron micrograph (Fig. 2a). The heterostructure was designed to form a very straight nanowire with the band-gap wavelength remaining below the laser wavelength to avoid direct optical absorption from the 1550-nm probe light. A WGM microsphere with a diameter of 40 $\mathrm{\mu m}$ was fabricated from a silica optical fiber by the standard discharge technique and contacted to a tapered optical fiber, which enables fiber access to the WGM using a telecom-wavelength light. The fiber-coupled microsphere was attached to a plate with glue and placed in a vacuum chamber ($\sim 10^{-5}$ Pa) together with the nanowire on the InP substrate. The linewidth of the WGM  is $\kappa /2\pi =$ 1.1 GHz, which corresponds to the quality factor $Q$ = $1.8 \times 10^5$ (Fig. 2b). The nanowire-containing InP substrate was put on a three-axis piezo-positioner, which enables an adjustment of the relative gap between the nanowires and microsphere along the $x$-axis, where we define the $x$-, $y$-, and $z$-axis such that the $x$-axis corresponds to the radial direction of the WGM as shown in Fig. 2c. Using the piezo-positioner, the microsphere was evanescently coupled to a single nanowire, which was located at the edge of the cleaved surface with the normal axis corresponding to [110].

A balanced homodyne interferometer was constructed to directly readout cavity phase modulation for high-sensitivity displacement measurement (see Fig. 2d). Light from an external diode laser was divided into a probe and local oscillator. The probe light was coupled to the microsphere via the optical tapered fiber inside the vacuum chamber. After it had passed through the optical cavity, the probe interfered with the local oscillator with a fixed power of 3 mW. The optomechanical signals were detected by a balanced photodetector. The radio-frequency part of the signal, i.e. signal in mechanical motion, was monitored with a spectrum analyzer, and the dc part of the signal was sent to a servo-controller to stabilize the optical phase of local oscillator via a piezo-loaded mirror. The optical phase of local oscillator was set on in-phase, i.e., only dispersive optomechanical readout was available. To achieve maximum efficiency, the on-resonance probe light was used in the following experiment.

Figure 3a shows the displacement noise power spectrum (at $x,y\approx 500$ nm, 0 nm, the definition of the position is given later) measured with an optical probe power of 3 $\mathrm{\mu W}$. This spectrum was obtained by calibrating the optical phase noise power spectrum with the application of an optical calibration tone with an electro-optic modulator (EOM) \cite{gorodetksy2010determination} and by taking into account the amount of the thermal motion and the nanowire's effective mass of 2.0 fg. It displays the thermal motion of the two orthogonalized mechanical modes of the  nanowire at 1.278 and 1.284 MHz hereafter referred to as mode 1 and mode 2, respectively. The linewidth  of the two modes is 283 and 222 Hz, respectively, which leads to the corresponding mechanical quality factors of $4.5 \times 10^3$ and $5.8 \times 10^3$. These two mode frequencies show good agreement with those of fundamental flexile modes numerically calculated by the finite element method. This reveals that the two spectra correspond to the nearly-degenerated orthogonalized modes of the standing nanowire, where the degeneracy was lifted due to the slight geometric asymmetry. Note that the displacement of the nanowire was measured via linear optomechanical coupling, where the output optical phase quadrature is proportional to the mechanical displacement in the bad cavity regime. The resultant linear optomechanical coupling constants, $g_\omega^{(1)}\equiv \frac{\partial\omega_\mathrm{cav}}{\partial x} x_\mathrm{zpf}$, are $2\pi$ $\times$ 28.5 and 23.9 Hz for the mode 1 and mode 2, respectively, where $\omega_\mathrm{cav}$ is an optical resonance frequency, $x_\mathrm{zpf}$ is a zero-point fluctuation, and $x$ is a mechanical displacement. The probe power of 3 $\mathrm{\mu W}$ results in a noise floor level (i.e. displacement sensitivity) of $2.4\times10^{-12}$ $\mathrm{m/\sqrt{Hz}}$. The noise floor level was further reduced as in the pump power was increased because the net linear optomechanical coupling is enhanced in proportion to the square of  the probe power (see Fig. 3b). The noise floor level of $8.2\times10^{-13}$ $\mathrm{m/\sqrt{Hz}}$ was achieved with the optical power of 15 $\mathrm{\mu W}$. This noise floor level is equivalent to the level of thermal motion at $2.8$ K. This displacement sensitivity implies that thermal motion can be detected at the liquid helium temperature, where rich quantum optical properties appear in heterostructured nanowires. The displacement sensitivity could be further improved to the standard quantum limit by optimizing the taper-sphere coupling (see Discussion).

The strength of optomechanical coupling depends on the vibration axis that is thus, in general, different between the two mechanical modes. In particular, the WGM optical field has a strongly anisotropic optical gradient where the radial direction of the microsphere ($x$ direction defined in our experimental coordinates) has a stronger gradient than the other directions (polar and azimuthal directions). Thus, the vibrating mode along the $x$ direction is better resolved than the orthogonal modes. On the other hand, the two-mode mechanical spectra in our experiment were almost equivalently observed via the displacement measurement as shown in Fig. 3a. This indicates that the initial configuration between the microsphere and nanowire has an almost $\pm 45$-degree angle between the radial direction of the WGM and the two orthogonalized vibration directions. To quantitatively evaluate this initial angle $\theta_0$, we measured two-mode thermomechanical spectra by sweeping the $y$ position (see Fig. 3c). Here, we define $y=0$, where the difference among the areas of power spectral density becomes zero. Being symmetric with respect to $y = 0$, the mechanical spectrum of mode 1 (2) dominantly appeared at the positive (negative) $y$ positions. The initial angle $\theta_0$ can be estimated by fitting the difference between the areas of power spectral density $\eta(y)\equiv P_1(y)- P_2(y)$, where $P_i(y)$ is the integrated power in the power spectral density in mode $i$. By fitting the experimental values with the theoretical ones, $\theta_0$ is estimated to be $44.9$ degrees (see Fig. 3d), where $\theta_0$ is extracted with the additional fitting parameter, $\phi_0$, which corresponds to the angle between $y$ axis and practical sweep direction. Note that such an initial configuration with $\theta\approx 45$ degrees originates in the fact that the optical microsphere was accessed at the edge of the cleaved surface with the normal axis corresponding to [110].

\subsection*{Control of mechanical responses via static optical forces}
Since the two orthogonalized mechanical modes are both coupled to the optical cavity mode, the optical force influences their frequency, linewidth, and the vibration axis. This can be understood from the following equations of motion:
\begin{align}
\ddot{X}_1+\Gamma_1\dot{X}_1+\Omega_1^2X_1=&F^\mathrm{th}_1+F_1^\mathrm{opt}(X_1,X_2),\\
\ddot{X}_2+\Gamma_2\dot{X}_2+\Omega_2^2X_2 =&F^\mathrm{th}_2+F_2^\mathrm{opt}(X_1,X_2),
\end{align}
where $X_i$, $\Gamma_i$, and $\Omega_i$ are the displacement, damping factor, and angular frequency of $i$th modes (we define $\Omega_2\geq \Omega_1$). The two modes are driven by the thermal Langevin force $F^\mathrm{th}_i$ under the optical force field $F_i^\mathrm{opt}(X_1,X_2)$ for the vibration direction of mode $i$. In the case of thermal motion with small mechanical displacements, we can approximate the optical force field as
\begin{align}
F_i^\mathrm{opt}(X_1,X_2)\approx F_i^0+g_{i1}(x)X_1+g_{i2}(x)X_2 \label{eq:Fopt}
\end{align}
where $F_i^0$ is a constant force to change the equilibrium points of vibration (thus, ignored in the following discussion), $g_{ij} (x)\equiv \partial_{X_{j}}F_i^\mathrm{opt}$ are the force gradient with a derivative along the $X_j$ direction. In this model, the optical force is assumed to be conservative. Note that the optical spring effect, which originates from the linear optomechanical coupling and often appears in a cavity optomechanical system by detuning the cavity \cite{aspelmeyer2014cavity}, does not appear under the current homodyne measurement with an on-resonance probe. Thus, the quadratic optomechanical coupling is the dominant source of this optical force \cite{doolin2014nonlinear}. The strength of this quadratic (dispersive) optomechanical coupling exponentially changes with the gap, i.e. $g_{ij}(x)$ is an exponent function for $x$. Therefore, this leads to the exponential frequency shift of the mechanical modes with the gap. Here, note that $g_{ij}$ also leads to the rotation of the vibration axis of the two mechanical modes when the off-diagonal component is a non-zero value (see Methods) \cite{de2017universal}. When we include this rotation effect, we can theoretically estimate the $x$ dependence of the two-mode spectra in the frequency domain (Fig. 4a). The theoretical results show an exponential shift of the two mode frequencies for $x$, which is in good agreement with the experimentally measured shift (Fig. 4b). The only difference between the experiments and theory appears in the linewidth. Experimentally, the linewidth broadens as the gap is reduced. This linewidth broadening is caused by the dissipative optomechanical coupling \cite{sawadsky2015observation}, which is not included in our theoretical model [Eq. (\ref{eq:Fopt})].

Now, one can find that the amount of frequency shift and the linewidth are not equivalent between the two mechanical modes (Figs. 4c and 4d). This difference is due to the rotation of the vibration axis by the optical force. It can be seen in the theoretical results by comparing the case including the rotation effect (color plots in Fig. 4a) and the case excluding it (solid curve in Fig.
4a). The optically induced rotation of the vibration axis pushes the frequency shift in mode 1, whereas it suppresses the frequency shift in mode 2. The angle between the vibration axis and the $x$-axis, $\theta$, can be quantitatively evaluated by taking into account the ratio of the frequency and the area of the displacement noise power spectrum between the two modes \cite{rossi2017vectorial}:
\begin{align}
\theta \equiv \tan^{-1}\left(\frac{\Delta\Omega_1}{\Delta\Omega_2}\sqrt{\frac{P_1}{P_2}}\right)
\end{align}
where $\Delta\Omega_i$ is the frequency shift in mode $i$. Figure 5a shows the gap ($x$) dependence of the angle $\theta$ at $y = 0$, which was extracted from Eq. (4). The experimental values (open circles) show good agreement with the theoretical values (solid curve, see Eq. (\ref{eq:theta}) in Methods), indicating rotation to $\theta = 18.9$ degrees at $x = 0$.

This rotation leads to the difference in the optomechanical coupling coefficients between the two modes, in which the coefficient of mode 1 is increased at the price of mode 2, especially at the small gap. The linear optomechanical coupling constant $g_\omega^{(1)}$, which contributes to the displacement sensitivity, was estimated by taking into account the linewidth broadening (see Methods). Apparently, $g_\omega^{(1)}$ in mode 1 gained several hundred hertz as decreasing the gap due to the rotation, whereas that in mode 2 converged (see Fig. 5b). Because this rotation aligns the vibration axis in mode 1 so that it becomes parallel to the $x$-axis, we can achieve optimized displacement sensitivity in mode 1 regardless of the initial angle. Such mode-selective coupling due to the rotation also appears in the quadratic optomechanical coupling coefficients. The quadratic optomechanical coupling coefficients are classified into a dispersive part, $g_\omega^{(2)}\equiv\frac{1}{2}\frac{\partial^2\omega_\mathrm{cav}}{\partial x^2} x^2_\mathrm{zpf}$, and a dissipative part, $g_\kappa^{(2)}\equiv \frac{1}{2}\frac{\partial^2\kappa}{\partial x^2} x^2_\mathrm{zpf}$. The dispersive (dissipative) part is estimated from the frequency shift (linewidth broadening) of the two modes (see Methods). The value of coefficients in mode 1 in the small gap region becomes about ten times larger than that in mode 2. Such mode-selective control of both mechanical frequency and linewidth could be extended to non-reciprocal operation \cite{de2018eigenmode} and quantum nondemolition measurement \cite{anetsberger2009near,li2012proposal,hauer2018phonon} via the quadratic optomechanical coupling.

In addition to the linear mechanical responses (i.e. frequency, linewidth, and vibration axes), the Duffing mechanical nonlinearity can be controlled by using the cavity-induced optical force. The Duffing nonlinearity often appears in strongly driven mechanical resonators as a result of the hardening or softening effect \cite{lifshitz2008nonlinear}. Such nonlinearity is externally uncontrollable because it is generally caused by the structural or material properties of resonators. The fourth-order optomechanical coupling, whose coefficient is given by $g_\omega^{(4)}\equiv \frac{1}{4!}\frac{\partial^4 \omega_\mathrm{cav}}{\partial x^4}x_\mathrm{zpf}^4$, with the on-resonance probe allows us to induce and control such Duffing nonlinearity in a free-standing structure like the present nanowire (see Fig. 6a). We performed the driven measurement with a lock-in detector under a strong excitation of the mechanical motion of the nanowire using a piezoelectric sheet set below the InP substrate (see Methods). Figure 6b and 6c show the mechanical response measured at two different gaps, $x \approx$ 100 and 500 nm, respectively. The symmetric mechanical spectra in the large gap $x\approx$ 500 nm reflect that the free-standing structure of nanowire does not show the Duffing nonlinearity (see Fig. 6b). On the other hand, the asymmetric mechanical spectra in the small gap $x\approx$ 100 nm with the same driving force indicate that the near-field optical force induces the Duffing nonlinearity via the fourth-order optomechanical coupling (see Fig. 6c). Note that the fourth-order optomechanical coupling coefficient under the on-resonance probe can give rise to only the softening effect ($g_{\omega}^{(4)}>0$). This positive sign is reasonable because $g_\omega^{(2)}>0$ as the same even-order coefficient due to the exponential profile of the optical fields.  Introducing an additional control light and setting its detuning to a finite value might make it possible to tunably adjust the softening and hardening effects like the optical spring effects can be \cite{aspelmeyer2014cavity}. Such an optically tunable Duffing nonlinearity could be used for sensing applications of nanomechanical resonators \cite{kumar2011bifurcation,chowdhury2020weak}.

Besides the Duffing nonlinearity, the near-field optomechanical coupling might induce richer mechanical nonlinearities in terms of the even-order mechanical nonlinearity and nonlinear mechanical damping \cite{eichler2011nonlinear,eichler2013symmetry}. The former can modulate the equilibrium point of mechanical motion in higher-odd-order optomechanical coupling (i.e. coefficient of $g_\omega^{(2n-1)}$ with $n\geq 2$), and the latter can enrich the dynamics of energy relaxation in the higher-order dissipative optomechanical coupling  (i.e. coefficient of $g_\kappa^{(n)}$ with $n\geq 3$). Note that introducing a control laser with finite detuning might make it possible to tune the sign of these nonlinear effects. Investigating the nonlinear mechanical properties in near-field cavity optomechanics might open the way to further functionalize the manipulation of nanowire mechanical systems.

\section*{Discussion}
The optomechanical performance achieved in our experiment could be further improved by optimizing the optical taper-cavity coupling with additional positioners. This optimization would increase the loaded optical Q factor, which is typically larger than $10^7$ in a silica microsphere of the current size. We estimate that improving the optical Q factor to $1.0 \times 10^7$ (i.e. $\kappa/2\pi$ = 20 MHz) could lead to displacement sensitivity of $2.9 \times 10^{-15}$ $\mathrm{m/\sqrt{Hz}}$, which corresponds to the standard quantum limit in this nanowire mechanical resonator, if a similar level of optomechanical coupling, $g_0/2\pi$ = 100 Hz, is obtained with the probe power of 16 nW. This implies that the Heisenberg-limit displacement measurement in the back-action-dominant regime would be available in this nanowire at room temperature \cite{schilling2016near}. Moreover, introducing the ultrahigh-Q microsphere ($Q>10^8$) \cite{ma2007radiation} would allow us to bring the system into the resolved-sideband regime ($\Omega>\kappa$), which would lead to rich optomechanics (e.g. the sideband cooling and heating) in semiconductor nanowires.

In principle, this near-field optomechanical approach does not limit the nanowire structure. Therefore, it can be extended to various types of nanowires with different sizes and inner structures, including arrayed nanowires \cite{huber1994nanowire,wang2006piezoelectric}. The extension to epitaxially engineered nanowires with an optical or electronic functionality would also be available. This would thus allow us to construct novel quantum hybrid cavity optomechanical systems based on, for example, nanowire field-effect transistors and quantum dots \cite{husain2003nanowire,solanki2011high,kim2019nanomechanical,tomita2020novel}, which could lead to highly efficient electro-optic transduction and highly sensitive charge detection through optomechanical coupling \cite{midolo2018nano}. The near-field cavity optomechanical coupling could also be demonstrated at liquid-helium temperatures by carefully designing the cryo-measurement setup. Such a cryogenic operation would open the way towards a novel hybrid quantum optoelectromechanical system, in which mechanical displacement (i.e. phonons) is resolved at a level close to the zero-point motion and a small number of electrons confined in the nanowire couple to a small number of photons via the cavity-induced optomechanical coupling. Moreover, coupling a cavity to a quantum-dot-embedded free-standing nanowire would lead to novel nanomechancial systems that show rich physics related to exciton-polariton dynamics \cite{restrepo2014single,kyriienko2014optomechanics}.

\section*{Methods}
\subsection*{Fabrication of InP/InAs heterostructured nanowire}
We prepared InP/InAs nanowires in a metal organic vapor phase epitaxy system \cite{zhang2019telecom}. We used bottom-up vapor-liquid-solid (VLS) growth for nanowire synthesis. Indium particles were used to catalyze nanowire growth in the self-catalyzed VLS mode. They were formed on heated InP (111)B substrate (360 ${}^\circ\mathrm{C}$) by introducing trimethylindium (TMIn) source material for 5 min at a flow rate of 3.03 $\mathrm{\mu}$mol/min \cite{zhang2012vertically}. We then reduced the temperature to 330 ${}^\circ\mathrm{C}$ and started nanowire growth by introducing TMIn and tertbutylarsenic (TBA) or tertbutylphosphine (TBP) simultaneously. The nanowire growth initiated from indium particles distributed on InP substrate. Owing to the epitaxial relationship between nanowires and substrates, the $\langle111\rangle$-oriented nanowires are vertically aligned on the substrates (Fig. 2a). By alternating source As and P materials, we added multiple InAs layers to InP nanowires. Each nanowire shown in Fig. 1a contains 300 InP/InAs units. For the InP segment growth, the flow rates of TMIn and TBP were 3.03 and 803.6 $\mathrm{\mu}$mol/min, respectively. For the InAs segment growth, the flow rates of TMIn and TBA were 3.03 and, 26.8 $\mathrm{\mu}$mol/min, respectively. The growth times for InP and InAs segments in each InP/InAs unit were 20 and 2 s, respectively. The thicknesses of the segments were 35$\pm$3 nm and 7.5$\pm$2 nm, respectively.

\subsection*{Estimation of initial angle $\theta_0$}
To quantitatively estimate the initial angle $\theta_0$ between the vibration direction and $x$ direction (the radial direction of the WGM), we measured the two-mode mechanical spectra by sweeping the $y$ position and calculated the difference in the integral of the power spectral density between the two modes for each $y$, $\eta (y)$. Note that the $x$ coordinate, which is defined so that $x=0$ on the sphere surface, was far enough ($x\approx 500$ nm) to avoid angle rotation due to the static optical force. By approximating that the two mechanical modes have the same effective masses and the same thermal occupation, the integrated power spectral density of each mode $P_i$ only reflects the linear optomechanical transduction for each $y$. Since the linear optomechanical transduction is proportional to the derivative of optical near-field potential along the vibration direction, $\eta (y)$ is modeled by
\begin{align}
\eta (y)\equiv&P_1(y)-P_2(y)\nonumber\\
& =\int_\mathrm{mode1}\mathrm{d}\omega S_{VV}(\omega,y)- \int_\mathrm{mode2}\mathrm{d}\omega S_{VV}(\omega,y)\nonumber\\
\approx& B(G_1(y)-G_2(y)) \label{eq:eta}
\end{align}
where $G_i(y)\equiv \partial U_\mathrm{opt}(x, y)/\partial X_i$ with $X_i$ the vibration coordinate in mode $i$, and $B$ an arbitrary constant. The vibration coordinate can be transformed to $X_1=x\cos \theta_0+y\sin\theta_0$ and $X_2=-x\sin \theta_0+y\cos\theta_0$ with the laboratory coordinate $(x,y)$ with the initial angle $\theta_0$. The optical near-field potential in the WGM of microsphere is given by
\begin{align}
U_\mathrm{opt}=&A|\psi^2_\mathrm{WGM}(x,y)|^2\nonumber\\
=&\left|j_l(k\sqrt{(x+R)^2+y^2}) P_m^l\left(\frac{y}{\sqrt{(x+R)^2+y^2}}\right)\right|^2 \label{eq:WGM}
\end{align}
where $\psi_\mathrm{WGM}(x,y)$ is the optical near-field of WGMs given by the product of a spherical Bessel function $j_l$ and associated Legendre function $P_m^l$ with the integers of $l$ and $m$, $R$ is the microsphere radius, and $k=2\pi/\lambda$ is the wavenumber of light with the resonance wavelength $\lambda$. Here, we fixed $l=m=220$ and $\lambda=1567$ nm from the boundary condition of WGM cavity \cite{matsko2005review}. By calculating the derivative of Eq. (\ref{eq:WGM}) along $x_i$ and substituting it into Eq. (\ref{eq:eta}), we achieved the theoretical model with free parameters $\theta_0$, $x$, and $y$. To fit this model into the experimental results, we assumed that $x\to x_0 +\tan \phi_0 y$ with $x_0$ fixed at $500$ nm and that $\phi_0$ is a tilted sweep angle to take into account of the imperfect configuration. As a result, we achieved $\theta_0= 44.9$ and $1.2$ degrees from the experimental data by fitting with the theoretical model.

\subsection*{Two-mode mechanical spectra with static optical forces}
From Eqs. (1) and (2), we analytically derives the two-mode mechanical spectra with Fourier transformation:
\begin{align}
\left(\begin{array}{cc}
\chi_1(\omega)-g_{11}(x)& -g_{12}(x)\\
-g_{21}(x)&\chi_2(\omega)-g_{22}
(x)\end{array}\right)
\left(\begin{array}{c}X_1\\X_2\end{array}\right)=\left(\begin{array}{c}
F_1^\mathrm{th}\\F_2^\mathrm{th}\end{array}\right)
\end{align} 
where $\chi_i(\omega)=\Omega_i^2-\omega^2+i\omega\Gamma_i$. Here, we assume that the static optical force makes an angle of $\pm45$ degrees in the vibration directions. This assumption of the initial angle reduces to $g_{ij}(x)=f_x(x)/2$, where $f_x(x)$ is the force gradient along the $x$ direction in the laboratory frame (shown in Fig. 2c). Note that we approximate that $f_x\equiv |\partial_x F_x|\gg |\partial_y F_x|, |\partial_x F_y|, |\partial_y F_y| \approx 0$ by taking into account the WGMs. The matrix is diagonalized as $\mathrm{diag}[\chi_+(\omega), \chi_-(\omega)]$ with
\begin{align}
\chi_\pm(\omega)=&\frac{1}{2}\Biggl[\chi_1(\omega)+\chi_2(\omega)-f_x(x)\nonumber\\
&\pm\sqrt{\left(\chi_1(\omega)-\chi_2(\omega)\right)^2+f_x(x)^2}\Biggr]\label{chipm}
\end{align}
Here, we assume that the initial angle between the vibration direction and force direction is $45$ degrees and regard the optical force gradient as an exponential function $A\exp\left(-x/x_\mathrm{ev}\right)$, $x_\mathrm{ev}\approx\lambda/(4\pi\sqrt{n_\mathrm{SiO_2}^2-1})$ with the optical wavelength $\lambda$ and refractive index of optical cavity $n_\mathrm{SiO_2}=1.44$. The constant $A=7\times 10^{-11}$ was chosen so that the amount of frequency shift almost corresponded to the experimental data. 

The two-mode mechanical spectra were experimentally achieved via the linear optomechanical coupling, whose strength also depends on the vibration direction. The linear optomechanical coupling constants, which are proportional to the optical force, are given by $F_x \cos\theta$ and $F_x \sin\theta$ where $F_x\approx \int\mathrm{d}x f_x=Ax_\mathrm{ev}\exp(-x/x_\mathrm{ev})$ and $\theta$ is the angle between the vibration direction and the dominant optical force in the radial direction of WGM cavity expressed by
\begin{align}
\theta(x)=\frac{\pi}{4}-\frac{\Omega_2^2-\Omega_1^2+\sqrt{f_x(x)^2+(\Omega_2^2-\Omega_1^2)^2}}{f_x(x)}\label{eq:theta}.
\end{align}
Note that the contribution of mechanical dissipation is ignored because the Q of the mechanical resonance is high enough with $\Omega_i\Gamma_i\ll \Omega_i^2$. The sum of the squares of amplitude $|\chi^{-1}_+(\omega)F_x\cos\theta|+|\chi^{-1}_-(\omega)F_x\sin\theta|$ are shown in Fig. 4a with an appropriate noise floor with the signal-noise ratio of $10^2$, which almost corresponds to the experimental one. Note that the non-rotating case described by the black dotted lines in Fig. 4a corresponds to the case of $g_{12}=g_{21}=0$.

\subsection*{Estimation of linear optomechanical coupling constant}
The linear optomechanical coupling constant was evaluated with a calibration method. The dispersive optomechanical coupling constant is given by
\begin{align}
g_\omega^{(1)}=&\sqrt{\frac{\int\mathrm{d}\omega S^\mathrm{mech}_{VV}(\omega)}{\int\mathrm{d}\omega S^\mathrm{ref}_{VV}(\omega)}}\frac{\Omega_\mathrm{ref}\beta}{\int\mathrm{d}\omega S_\mathrm{xx}(\omega)}
\end{align} 
where $S^\mathrm{mech}_{VV}(\omega)$ and $S^\mathrm{ref}_{VV}(\omega)$ are experimentally measured power spectral densities with the unit of $\mathrm{V^2}/\mathrm{Hz}$, and $\Omega_\mathrm{ref}$ and $\beta$ are reference frequency and phase modulation rate, respectively \cite{gorodetksy2010determination}. In the case of no damping in a mechanical resonator (pure thermal motion), $\int\mathrm{d}\omega S_\mathrm{xx}(\omega)=4 n_M$ with the phonon number $n_M$. However, if there is an additional damping, $\int\mathrm{d}\omega S_\mathrm{xx}(\omega)$ should be corrected as $4 n_M \Gamma/\Gamma_0$, where $\Gamma$ is measured linewidth due to the additional damping from the quadratic optomechanical coupling and $\Gamma_0$ is an intrinsic linewidth of mechanical resonator. As a result, we estimated the linear optomechanical coupling constant by
\begin{align}
g_\omega^{(1)}=&\sqrt{\frac{\int\mathrm{d}\omega S^\mathrm{mech}_{VV}(\omega)}{\int\mathrm{d}\omega S^\mathrm{ref}_{VV}(\omega)}}\frac{\Omega_\mathrm{ref}\beta}{4 n_M}\frac{\Gamma_0}{\Gamma}\label{corr_g}.
\end{align} 
Importantly, the corrected optomechanical coupling with Eq. (\ref{corr_g}) shows an exponential dependence on the gap distance, which is predicted in theory, whereas the noncorrected coupling does not.

\subsection*{Estimation of static quadratic optomechanical coupling constant}
The dispersive and dissipative quadratic optomechanical coupling constants, denoted by $g^{(2),i}_\omega$ and $g^{(2),i}_\kappa$, respectively, for the orthogonalized mechanical mode $i$ are defined by
\begin{align}
g^{(2),i}_\omega=&\frac{1}{2}\frac{\partial^2 \omega_\mathrm{cav}}{\partial X_i^2}x_{\mathrm{zpf},i}^2=\frac{1}{n_\mathrm{cav}}\mathrm{Re}\left[\frac{\partial F_{X_i}}{\partial X_i}\right]\\
g^{(2),i}_\kappa=&\frac{1}{2}\frac{\partial^2 \kappa}{\partial X_i^2}x_{\mathrm{zpf},i}^2=\frac{1}{n_\mathrm{cav}}\mathrm{Im}\left[\frac{\partial F_{X_i}}{\partial X_i}\right]
\end{align}
where $\omega$ and $\kappa$ are the cavity frequency and linewidth modified by the mechanical displacement $X_i$ of mode $i$, $x_{\mathrm{zpf},i}$ are zero-point fluctuations of mode $i$. The cavity photon number $n_\mathrm{cav}$ with zero-detuned probe is given by
\begin{align}
n_\mathrm{cav}=\frac{4\eta}{\kappa}\frac{P_\mathrm{in}}{\hbar\omega_0}
\end{align}
where $P_\mathrm{in}$ is the input probe power at the tapered fiber, $\hbar$ is Planck's constant over $2\pi$, $\omega_0$ is the cavity resonance frequency, and $\eta\equiv \kappa_\mathrm{in}/\kappa$ with taper-cavity coupling constant $\kappa_\mathrm{in}$ determines the coupling ratio at which the critical coupling condition corresponds to 1/2. In our experiment, $n_\mathrm{cav}$ was about $4.0\times 10^3$ with the input optical power of $3.0$ $\mathrm{\mu} W$. Although these expressions are equivalent to the ones in previous work \cite{doolin2014nonlinear}, we modified them to estimate the coupling constants with the experimental observables by taking into account the intermodal coupling.

Here, we regard the static optical force as a complex force where real part shows a conservative portion inducing a frequency shift, and its imaginary part shows a non-conservative one inducing linewidth broadening. 
To estimate the static quadratic optomechanical coupling constant from the frequency shift and linewidth broadening, Eq. (\ref{chipm}) is recalculated for an arbitrary rotated angle $\theta$ (i.e., $g_{11}(x)=f_x(x)\cos^2\theta, g_{12}(x)=g_{21}(x)=f_x(x)\cos\theta\sin\theta, g_{22}(x)=f_x(x)\sin^2\theta$) as follows:
\begin{align}
&\chi_\pm(\omega)=\frac{1}{2}\Biggl[\chi_1(\omega)+\chi_2(\omega)-f_x(x)\nonumber\\
&\pm\sqrt{\left(\chi_1(\omega)-\chi_2(\omega)\right)^2+f_x(x)^2-2\left(\chi_1(\omega)-\chi_2(\omega)\right)\cos2\theta}\Biggr].\label{chipm2}
\end{align}
The force gradient is decomposed to real and imaginary parts as $f_x(x)=f_R(x)+if_I(x)$, where each part can be calculated from Eq. (\ref{chipm2}) as  
\begin{align}
f_R(x)=-(\Delta\Omega_1+\Delta\Omega_2),\hspace{10pt}f_I(x)=-(\Delta\Gamma_1+\Delta\Gamma_2)
\end{align}
where
\begin{align}
\Delta\Omega_{1,2}=&\mathrm{Re}\left[\chi_{+,-}(\omega)-\left.\chi_{+,-}(\omega)\right|_{f_x\to0}\right],\\
\Delta\Gamma_{1,2}=&\mathrm{Im}\left[\chi_{+,-}(\omega)-\left.\chi_{+,-}(\omega)\right|_{f_x\to0}\right]. 
\end{align}
By transforming to the coordinate to $(X_1, X_2)$ (i.e. $f_{X_1}e\equiv \partial_{X_1}F_{X_1}\approx f_X\cos^2 \theta, f_{X_2}\equiv \partial_{X_2}F_{X_2}=f_X\sin^2 \theta$), we can represent the quadratic optomechanical coupling constant with experimentally measured parameters $\Delta\Omega_i$, $\Delta \Gamma_i$, and $\theta$, as follows:
\begin{align}
g^{(2),1}_\omega=&-\frac{\Delta\Omega_1+\Delta\Omega_2}{n_\mathrm{cav}}\cos^2\theta,\hspace{5pt}g^{(2),2}_\omega=-\frac{\Delta\Omega_1+\Delta\Omega_2}{n_\mathrm{cav}}\sin^2\theta,\\
g^{(2),1}_\kappa=&-\frac{\Delta\Gamma_1+\Delta\Gamma_2}{n_\mathrm{cav}}\cos^2\theta,\hspace{5pt}g^{(2),2}_\omega=-\frac{\Delta\Gamma_1+\Delta\Gamma_2}{n_\mathrm{cav}}\sin^2\theta,
\end{align}
which were estimated with the experimentally observed $\Delta\Omega_i$, $\Delta \Gamma_i$, and $\theta$ for each $x$ in Fig 5c and 5d.

\subsection*{Driven lock-in measurement for observing Duffing nonlinearity}
To confirm that the Duffing nonlinearity is induced by the near-field optical gradient, we performed the driven lock-in measurement as follows. The nanowire mechanical amplitude was electrically driven with a radio frequency drive signal via a piezoelectric sheet attached to the substrate. A lock-in amplifier was used to generate the rf drive signal and to measure the mechanical amplitude from the output of the balanced photoreceiver (shown in Fig. 2d). The two-mode mechanical spectra (shown in Fig. 6b and 6c) were acquired by sweeping the frequency of the drive signal from 1.28 to 1.29 MHz.

\section*{Acknowledgements}
We thank Samer Houri and Ryuichi Ohta for fruitful comments. This work was partly supported by JSPS KAKENHI (No. JP15H05869 and
No. JP15K21727).

\section*{Contributions}
T.T. and G.Z. designed and fabricated the semiconductor heterostructured nanowires. M.A. and H.O. performed the optomechanical measurements. M.A. analyzed experimental data with help from H.Y. and H.O.. M.A. and H.O. wrote the paper based on the discussion with all the authors. H.O. planned the project.

\begin{figure*}[htb]
\includegraphics[width=16.2cm]{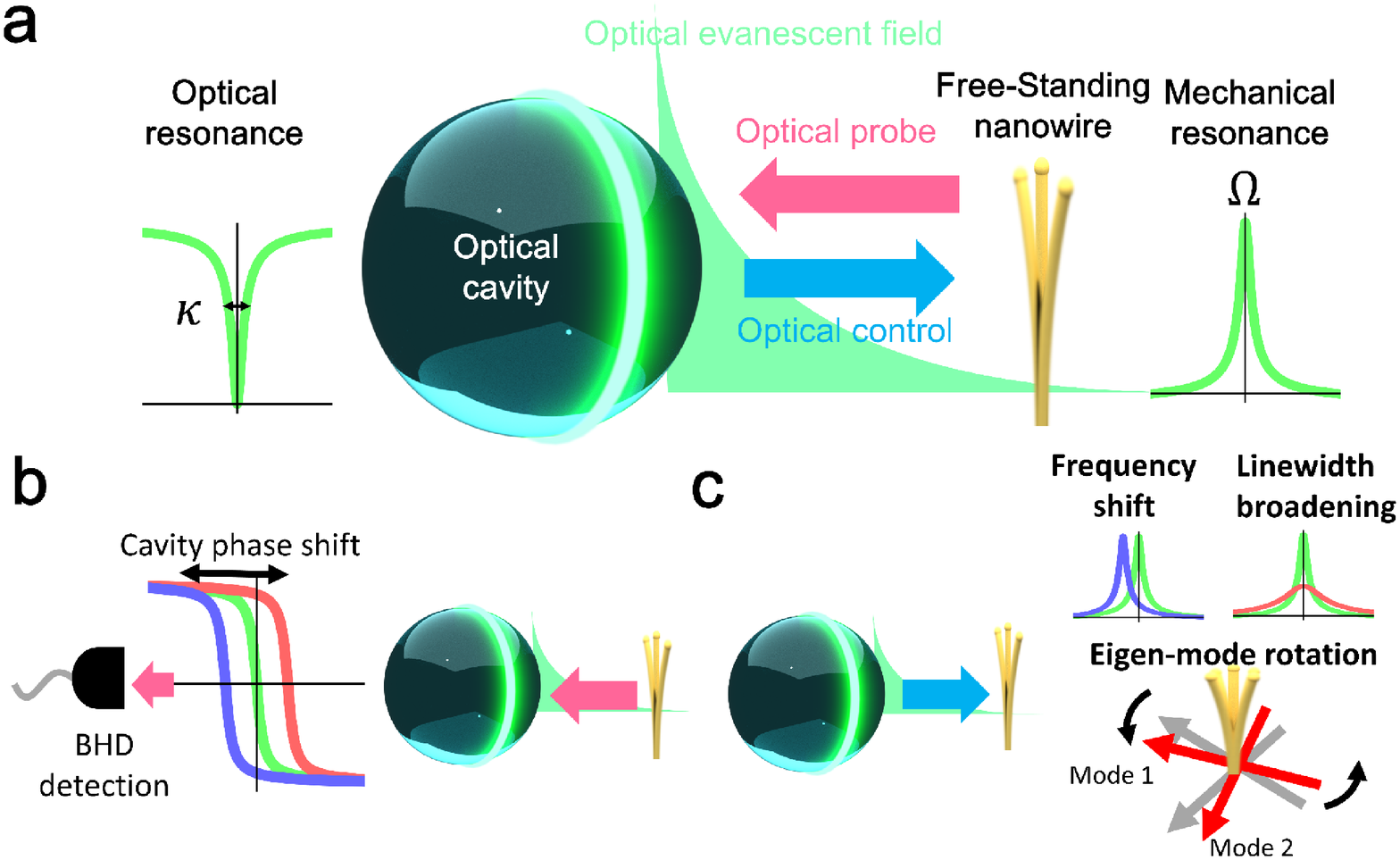}
\caption{\label{fig1} {\bf Conceptual illustration of near-field cavity optomechanics with a free-standing nanowire and a WGM microsphere}. {\bf a} Near-field cavity optomechanical coupling between the nanowire and microsphere via optical evanescent field. {\bf b} Optical probe of mechanical displacement via cavity phase modulation with balanced homodyne interferometry. {\bf c} Optical control of mechanical response with frequency shift, linewidth broadening, and eigen-mode rotation via static optical force.}
\end{figure*}
\begin{figure*}[htb]
\includegraphics[width=16.2cm]{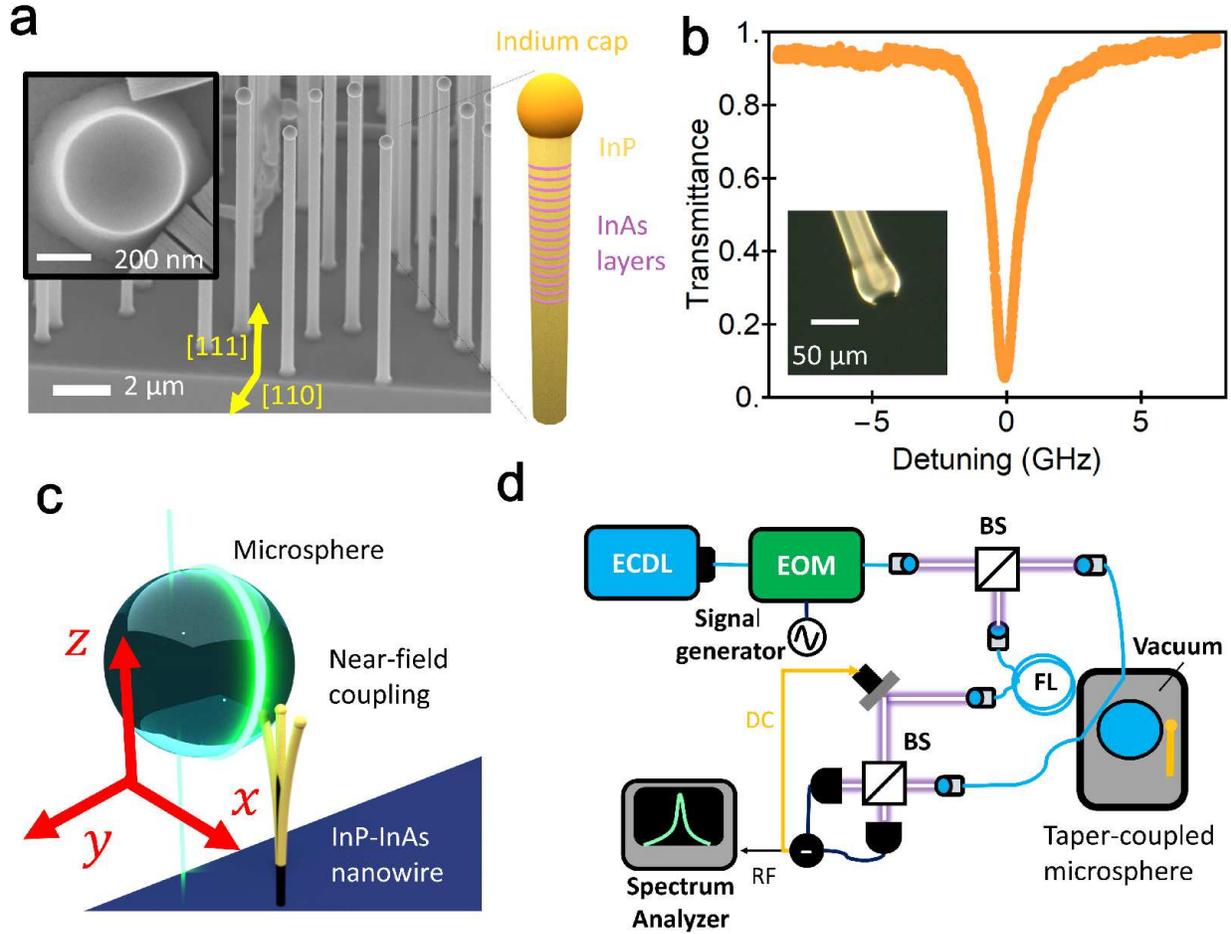}
\caption{\label{fig2} {\bf Device and setup for near-field cavity optomechanics}. {\bf a} Scanning microscope image of as-grown InP/InAs heterostructured nanowires and a schematic of the internal structure of a nanowire. The crystal orientation for growth direction [111] and direction of cleaved surface [110] are represented by yellow vectors. The inset shows a cross-sectional image of a nanowire. {\bf b} Transmission spectrum of the optical cavity with a microscope image of it (inset). {\bf c} Illustration of the configuration of the microsphere and nanowire with coordinates defined such that the radial direction of the WGM corresponds to $x$ axis. {\bf d} Schematic of optical measurement setup with a fiber-based balanced homodyne interferometer. ECDL: external cavity diode laser. EOM: electro-optic modulator. BS: beam splitter. FL: fiber loop. }
\end{figure*}
\begin{figure*}[htb]
\includegraphics[width=16.2cm]{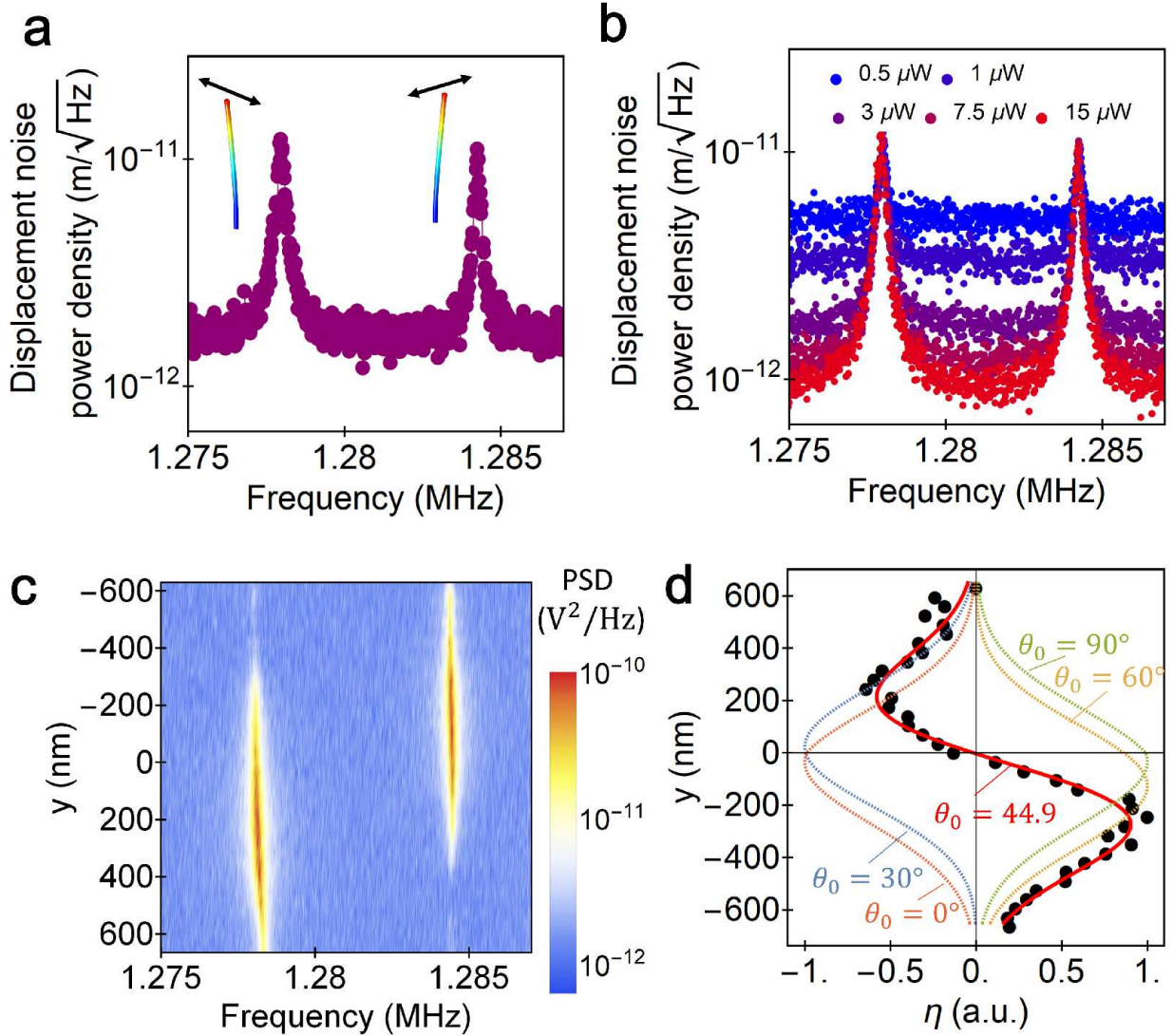}
\caption{\label{fig3} {\bf Characterization of two-mode mechanical spectra and their sensitivity in thermal motion by balanced homodyne interferometry}. {\bf a} Two-mode power spectral density of thermal motion observed by balanced homodyne interferometry with an optical probe power of 3 $\mathrm{\mu W}$. The insets show the fundamental flexural mode calculated by the finite element method. {\bf b} Two-mode power spectral density of thermal motion with different optical probe powers. The lowest noise floor level --$S_\mathrm{floor}=8.2\times10^{-13} \mathrm{m/\sqrt{Hz}}$-- was achieved with the optical probe power of $15$ $\mathrm{\mu W}$. {\bf c} Two-mode power spectral density of thermal motion with respect to $y$ coordinate.  {\bf d} Normalized difference between integral of power spectral density for the modes $\eta(y)$ with respect to $y$ coordinate. The black dots show the experimental data, and the red solid curve show the fitted curve with $\theta_0=44.9$ degrees and an tilted sweep angle $\phi_0=1.2$ degrees. $\eta$ has completely different profiles with respect to the initial angle $\theta_0$ between the $x$ axis and the vibrating direction of a mechanical mode in the nanowire (curves with $\theta_0={0, 30, 60}$, and 90 degrees are shown with $\phi_0=1.2$ deg).}
\end{figure*}
\begin{figure*}[htb]
\includegraphics[width=16.2cm]{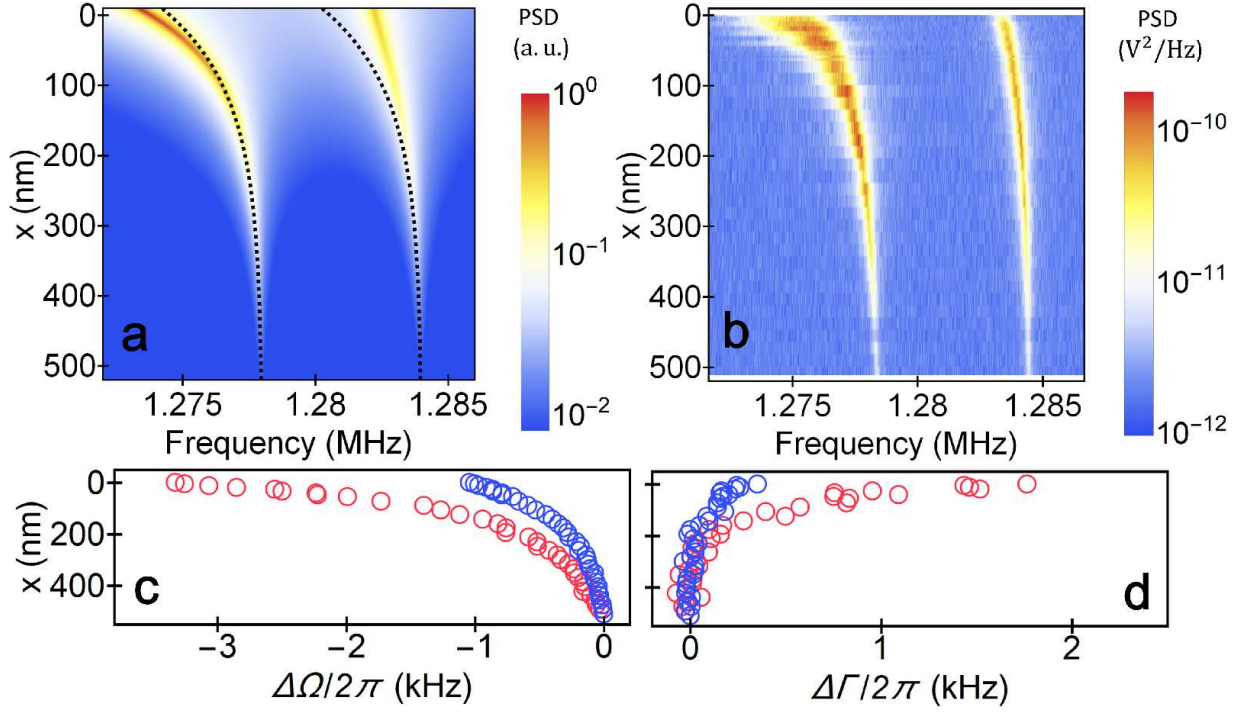}
\caption{\label{fig4} {\bf Two-mode mechanical responses with static optical forces}. {\bf a} Theoretically predicted two-mode mechanical spectra with optical force field in terms of gap $x$. The black dashed lines show the two-mode frequencies without rotation of eigen-modes. {\bf b} Experimentally observed two-mode power spectral density with respect to $x$ coordinate. {\bf c} and {\bf d} Mechanical frequencies and linewidths with respect to $x$ coordinate. The red and blue circle corresponds to them in mode 1 (lower-frequency mode) and mode 2 (higher-frequency mode), respectively.}
\end{figure*}

\begin{figure*}[htb]
\includegraphics[width=16.2cm]{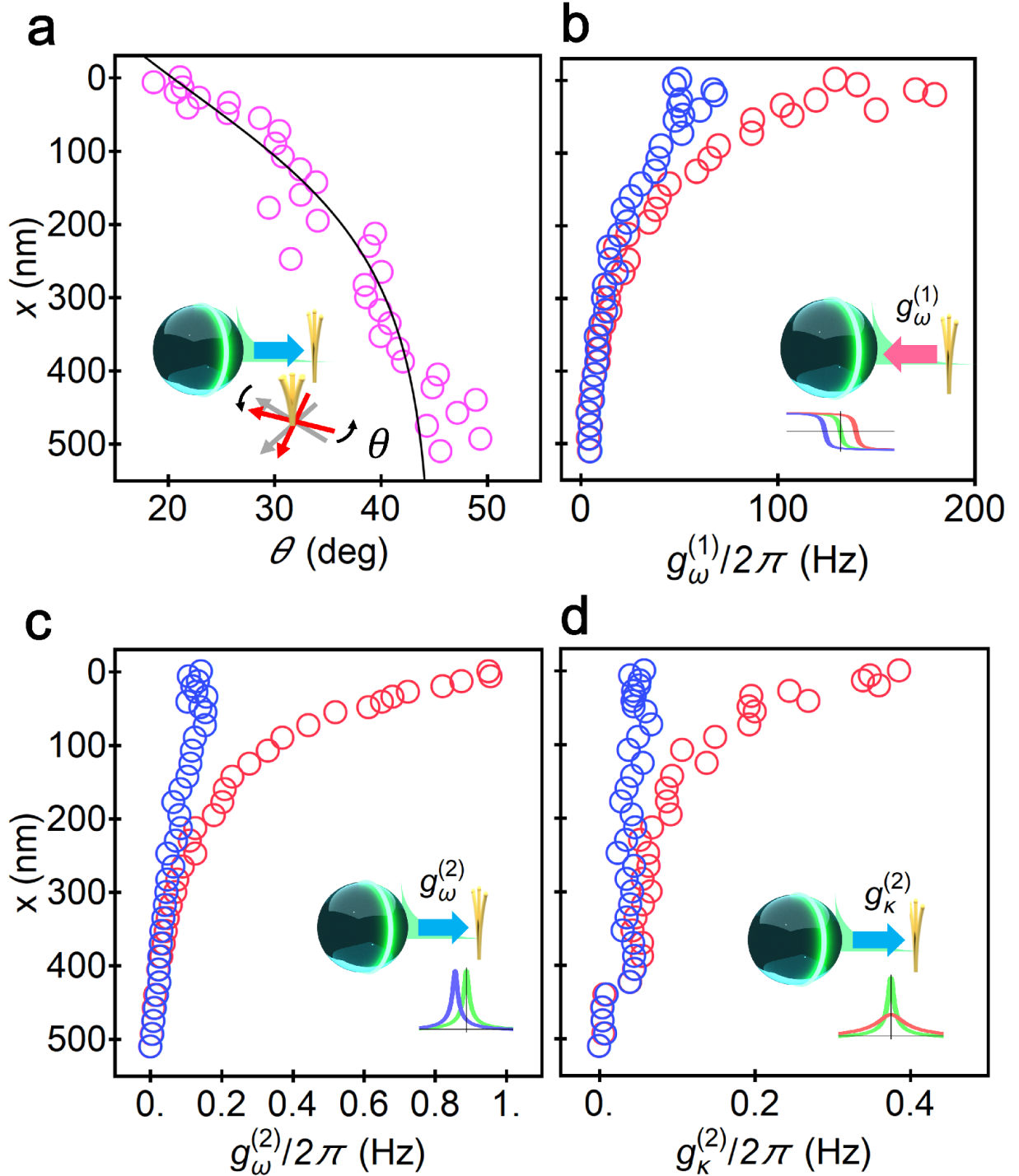}
\caption{\label{fig5} {\bf Eigen-mode rotation angle and optomechanical coupling constants}. {\bf a} Angle between $x$ axis and a vibrating mode in the nanowire. The purple circles are the experimentally observed rotation angles, and the black solid line shows data fitted with a theoretical model. {\bf b}, {\bf c} and {\bf d} Linear dispersive, quadratic dispersive, and quadratic dissipative optomechanical coupling constants estimated from net phase modulation, frequency shift, and linewidth broadening, respectively. The insets are conceptual illustrations of optomechanical coupling with each coupling constant.}
\end{figure*}
\begin{figure*}[htb]
\includegraphics[width=16.2cm]{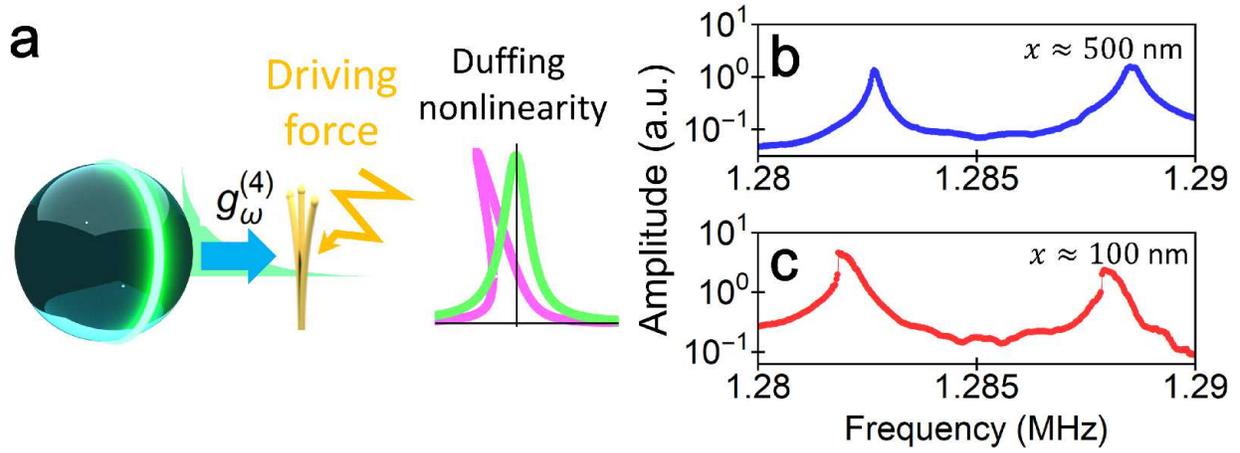}
\caption{\label{fig6} {\bf Optically-induced Duffing nonlinearity}. {\bf a} Conceptual illustration of optically-induced Duffing nonlinearity via the static fourth-order optomechanical coupling. Two-mode mechanical amplitude achieved in the driven lock-in measurement at $x\approx 100$ nm  {\bf b} and $x\approx 500$ nm  {\bf c}.}
\end{figure*}

\end{document}